\documentclass[prl,twocolumn,
superscriptaddress,showpacs,
]{revtex4}
\usepackage{amsmath}
\usepackage{graphicx}
\usepackage{dcolumn}
\usepackage{bm}
\usepackage{amssymb}
\usepackage{epsfig}

\def \beq {\begin{equation}}
\def \eeq {\end{equation}}
\def \ba {\begin{eqnarray}}
\def \ea {\end{eqnarray}}
\newcommand{\ra}{\rangle}

\def\ket#1{\left| #1\right>}

\def\lsim{\mathrel{\rlap{\lower4pt\hbox{\hskip1pt$\sim$}}
    \raise1pt\hbox{$<$}}}                
\def\gsim{\mathrel{\rlap{\lower4pt\hbox{\hskip1pt$\sim$}}
    \raise1pt\hbox{$>$}}}                

\begin{document}

\title{Fault-tolerant Quantum Communication with Minimal Physical Requirements}  


\author{L. Childress}
\affiliation{Department of Physics, Harvard University, Cambridge,
Massachusetts, 02138}

\author{J. M. Taylor}
\affiliation{Department of Physics, Harvard University, Cambridge, Massachusetts, 02138}

\author{A. S. S\o rensen}
\affiliation{Department of Physics, Harvard University, Cambridge, Massachusetts, 02138}
\affiliation{ITAMP, Harvard-Smithsonian Center for Astrophysics,
Cambridge, Massachusetts, 02138}
\affiliation{The Niels Bohr Institute, University of
Copenhagen, DK-2100 Copenhagen \O, Denmark}
\author{M. D. Lukin}
\affiliation{Department of Physics, Harvard University, Cambridge, Massachusetts, 02138}
\affiliation{ITAMP, Harvard-Smithsonian Center for Astrophysics,
Cambridge, Massachusetts, 02138}

\date{\today}
\begin{abstract}
  We describe a novel protocol for a quantum repeater which enables long
  distance quantum communication through realistic, lossy photonic
  channels. Contrary to previous proposals, our protocol incorporates
  active purification of arbitrary errors at each step of the protocol
  using only two qubits at each repeater station.  Because of these minimal
  physical requirements, the present protocol can be realized in simple
  physical systems such as solid-state single photon emitters. As an example, we show how nitrogen vacancy color centers in diamond can be
  used to implement the protocol, using the nuclear and electronic
  spin to form the two qubits.

\end{abstract}

\maketitle

Quantum communication holds promise for transmitting secure messages
via quantum cryptography, and for distributing quantum
information~\cite{gisin02}.  However, attenuation in optical fibers
fundamentally limits the range of direct quantum communication
techniques~\cite{brassard00}. If a photon is injected into an optical
fiber the probability to retrieve the photon after a distance $L$
decreases exponentially making transmission very impractical for long distances.
In principle, photon losses can be overcome by introducing
intermediate quantum nodes and utilizing a so-called quantum repeater
protocol~\cite{Briegel98}.  A repeater creates entanglement over long
distances by building a backbone of entangled pairs between
closely-spaced nodes.  Performing an entanglement swap at each
intermediate node~\cite{zukowski93} leaves the outer two nodes
entangled, and this long-distance entanglement can be used to teleport
quantum information~\cite{bennett93,bouwmeester97} or transmit secret
messages via quantum key distribution~\cite{ekert91}.  Even though
quantum operations are subject to errors, by incorporating
entanglement purification~\cite{Bennett96,Deutsch96} at each step, one
can create long-distance high-fidelity entangled pairs in a time that
scales polynomially with distance~\cite{Briegel98}.
 
In practice, current long-distance quantum communication schemes
remain challenging to implement in the laboratory.  For example, while
approaches based on photon storage in atomic ensembles~\cite{duan01}
are being explored, and these implementations are capable of
correcting errors caused by photon losses, they offer no protection
against more general errors such as those due to dynamical phase
fluctuations.  Other approaches attempt to create an interface between
light and single quantum bits (qubits)~\cite{blinov04,vanenk98}.  In
order for these schemes to be fully fault-tolerant, existing theories
\cite{Dur99} require that each node must contain a small quantum
computer whose size increases logarithmically with the communication
distance; the construction of such quantum computers represents a
considerable challenge.  In this Letter, we present a protocol for a
fully fault-tolerant quantum repeater in which each node is formed by
a two qubit quantum computer. We thereby avoid the increase in the
number of qubits required by previous protocols, which substantially
simplifies the experimental realization of quantum repeaters.

In addition to presenting an algorithm for resource-efficient
entanglement propagation and purification, we also present a physical
system in which it can be implemented.  The general protocol
is relevant to a variety of systems, including trapped atoms in a
cavity~\cite{mckeever04} or trapped ions~\cite{blinov04}.  The reduced physical requirements facilitate  development of a scheme which is,
to our knowledge, the first realistic proposal for the construction of
a quantum repeater in a solid state environment. 
 In particular,
we describe how the repeater nodes may constructed from single photon
emitters in solid state systems by using the nuclear spin degree of
freedom to store quantum information while the electron spin is used
for communication with neighboring nodes. This may be accomplished,
e.g., in nitrogen-vacancy (NV) centers in diamond
~\cite{weinfurter00,beveratos02}, single quantum dots \cite{santori02,
  bracker04, taylor03}, etc.  We start with the details of our
protocol for a fault-tolerant quantum repeater.  Then, a specific
implementation using NV centers in diamond is developed to demonstrate
the scheme.

  \begin{figure}\vspace{.0in}
\centerline {
\includegraphics[width=3in]{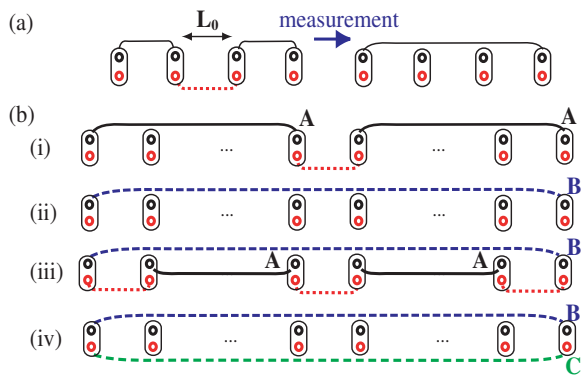}
}
\vspace{.0in}\caption{Protocol for fault-tolerant quantum
communication.   Each node is represented by an oval containing two qubits (circles).  Entangled states are represented by a solid or dashed line between the entangled qubits.
 (a) Entanglement connection.  
 (b) Nested entanglement purification.   Dots indicate an arbitrary number of
nodes.  
} 
\label{repeater}\end{figure}

A protocol for a simple repeater is presented in Fig. \ref{repeater}
(a). The total communication channel is divided into small segments of
length $L_0$ by a set of quantum nodes, each containing a two qubit
quantum computer.  Initially, two qubits in neighboring nodes are
prepared in an entangled state
$|\Psi_-\rangle=(|01\ra-|10\ra)/\sqrt{2}$, where $|0\ra$ and $|1\ra$
are the two states of the qubits [solid line between upper qubits in
the first and second node in Fig. \ref{repeater} (a)].  As detailed
below, we envision that such entangled states can be prepared
probabilistically between distant nodes with state-selective light
scattering.  Because the optical fibers connecting the two nodes are
lossy, this step is necessarily probabilistic, and has to be repeated
until successful.  Simultaneously, an entangled state is prepared
between two nearby nodes [solid line between upper qubits in the third
and fourth node in Fig.~\ref{repeater}(a)].  Once all entanglement
steps succeed the entangled pairs are mapped into storage states
(i.e., nuclear spins) and the remaining qubits in the central nodes
are entangled in a similar fashion [dotted line between lower qubits
in Fig. \ref{repeater} (a)].  Finally, an entanglement swap
\cite{zukowski93} is performed at each node, which teleports the
entanglement between the nodes so that eventually the outer two qubits
are entangled [Figure~\ref{repeater}(a), right].  Since entanglement
is generated only over a fixed distance $L_0$, this procedure avoids
the exponentially small probability for a photon to travel the full
length of the channel and thus allows the construction of long
distance entanglement from short range entanglement---provided that
there are no errors.

A single error in the chain will destroy the final state, making the
fidelity decay exponentially with distance.  To extend entanglement to
long distances in the presence of errors, active purification is
required at each level of the repeater scheme.  According to a standard
purification protocol ~\cite{Deutsch96},  we prepare two entangled pairs between two nodes.  Local two-qubit operations at each node are followed by measurement of one qubit at each node.   Conditioned on a successful outcome
of the measurement, this procedure yields an entangled pair of higher
fidelity between the remaining entangled qubits in the two nodes.  In Fig.~1(b) we present a protocol
for incorporating entanglement purification into a two-qubit repeater
scheme.  For clarity, we distinguish three types of entangled pairs A,
B, and C, labeled according to their purity.  A-pairs are fully
purified high fidelity pairs ready to be used in the next step of the
protocol, B-pairs are intermediate pairs, which are being purified to
A-pairs, and C-pairs are the lowest quality pairs which are used to
purify the B pairs.

The argument proceeds inductively: We assume that we have a method to
create and purify A pairs over distances up to $L_n = n L_0$ using
only two qubits per node and show that we can use these to generate
and purify A pairs over a distance $L_{2n+1} = (2n+1)L_0$.  It is
fairly straightforward to construct a B-pair over distance $L_{2n+1}$
by creating two purified A-pairs over the distance $L_n$, and
connecting them via an entangled pair between the central nodes (see
Fig 1(b)(i-ii)).  Previous schemes \cite{Briegel98, Dur99} have
constructed a C-pair in the same manner---at the cost of requiring an
extra qubit in the outermost nodes.  Instead, we employ the unused
nearest-neighbor nodes, creating two A-pairs and three short range
pairs, as shown in Fig.~\ref{repeater} (b)(iii).  Performing an
entanglement swap at the central and nearest neighbor nodes creates a
C pair over the distance $L_{2n+1}$ which can be used to purify the B
pair, see Fig.~\ref{repeater} (b)(iv).  We then perform the
purification protocol \cite{Deutsch96}, which, if successful, results
in a pair B-pair with higher fidelity.  The generation of C-pairs and
purification may then be repeated.  After generating C pairs for $m$
consecutive successful purification steps, (a technique sometimes
referred to as ``entanglement pumping"), the stored pair becomes a
purified A-pair over the full distance $L_{2n+1}$.  We remark that
this procedure is most efficient when $n \sim 2^k$ for integer $k$.

The fidelity obtained at the end of this nested purification
procedure, $F(m, L/L_0, F_0, p, \eta)$, depends on the number of
purification steps $m$, the total number of nodes $L/L_0$, the
initial fidelity $F_0$ between adjacent nodes, and the reliability of
measurements $\eta\leq1$ and local two-qubit operations $p\leq1$
required for entanglement purification and
connection.  There is little theoretical insight gained from the mathematical form of $F$ (see Refs.~\cite{Dur99,andersXX}) but it is easily evaluated numerically. In Fig. 2 we show the result of a numerical investigation of the
protocol in the presence of errors. In this analysis we assume that
the qubits do not decohere 
significantly over the communication time.
In Fig.~2 (a)  we show the fidelity as a function of distance. The curves show
the fidelity obtained by using three purification steps $m=3$. As seen
in the figure the fidelity saturates and only shows a very limited
decrease in fidelity with distance, demonstrating the ability of the protocol to
correct  error and the applicability of the protocol for long distance quantum
communication. As shown
in Fig. 2 (b),  the required time scales polynomially with distance.

\begin{figure}\vspace{.0in}
\centerline {
\includegraphics[width=3.5in]{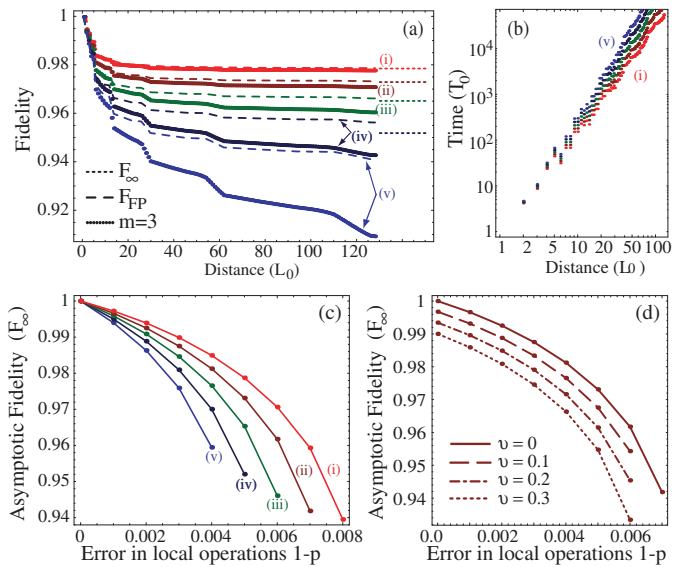}
}
\vspace{.0in}\caption{ 
(a) Fidelity scaling with distance.   Points show results using 3
    purification steps at each nesting level;  dashed lines show the
    fixed point $F_{FP}$ at each distance; dotted lines indicate the 
asymptotic fidelity $F_{\infty}$.   For (a) and (b), measurements and
local two-qubit operations $\eta=p$ contain $0.5\%$ errors.  For (a),
(b), and 
(c), the initial fidelity $F_0$ is (i) 100\% (ii) 99\% (iii) 98\% (iv)
97\% (v) 96\% with phase errors only.   (b) Time scaling with distance
for m=3, given in units of $T_0 \gg L_0/c$, 
the time required to
generate entanglement between nearest neighbors,  and $L_0$, the
distance between nearest neighbors. 
(c) Long-distance asymptote dependence on operation and measurement errors.  (d) Long-distance asymptote dependence on error
    type 
    ($F_0 = 0.99$,  
    $\upsilon = 0, 0.1, 0.2, 0.3$).}  
\label{asymp}\end{figure}

To determine the tolerance to the initial fidelity $F_0$ and to errors
in gates  $1-p$ and measurements
$1-\eta$, it is useful to consider the limiting case of many
purification steps and large distances.   
As the number of purification steps increases  $m\rightarrow \infty$,
the fidelity at a given distance $L$ grows, eventually saturating at 
 a fixed point 
 \beq
 F\rightarrow F _{FP}(L, F_0, p, \eta)\ .
 \eeq 
Additional purification steps yield no further benefit at the fixed point, because
the increase in fidelity they offer is cancelled by the likelihood of errors in the purification
procedure~\cite{Dur99}.   Moreover, as $L$ increases,  the fidelity
may 
approach an asymptotic value 
\beq
F_{FP}\rightarrow F_{\infty}(F_0, p,
\eta)\ ,
\eeq
 which is independent of distance~\cite{andersXX}.  For
comparison these quantities are also shown in Fig. \ref{asymp} (a). 
The asymptotic fidelity $F_\infty$ in Fig.~2 (c)  shows  
that our scheme will operate in the presence of $1-p \lsim 1\%$ errors
in local operations and percent-level phase errors in initial
entanglement fidelity.  For the specific physical system presented
below, the most likely error in entanglement generation between neighboring nodes results in an incoherent 
admixture of the state $|\Psi_+\ra=(01\ra+|10\ra)$, which we refer to as a phase error. 
Above we have assumed that only this type of error matters, but other types are
in principle possible.  In Fig.~2d we account for arbitrary errors in the initial entanglement by allowing  incoherent admixtures of the other two Bell states $\ket{\Phi_{\pm}} = (\ket{00}\pm
\ket{11})/\sqrt{2}$, each with weight $\upsilon(1-F_0)$~\cite{Deutsch96} (the weight of phase errors is thus $(1-2\upsilon)(1-F_0)$).     
Although the protocol we use is most effective for purifying phase
errors, Fig.~2d indicates that  it also tolerates arbitrary errors. 

For the implementations discussed below, the overall time scale is set
by the classical communication time between nodes, and the fidelity is
limited by the photon emission probability $P_{\rm{em}}$ from each
node. 
As an example,
using a high photon 
collection efficiency, a photon loss rate of $\sim 0.2$ dB/km,
spacing $L_0 \sim 20$ km, 
an initial fidelity $F_0$ set by 
an emission probability $P_{\rm{em}} \sim 8\%$ (see Eq.~\ref{fidelity} below), 
local errors $\eta = p = 0.5\%$, and just one purification step at
each nesting level, our scheme could potentially produce entangled
pairs with fidelity $F\sim 0.8$ sufficient to violate Bell's
inequalities over 1000 km in a few seconds.    
Moreover, 
the bit-rate could likely be significantly improved by employing
optimal control theory to tailor the details of the repeater protocol
to the parameters of a desired implementation.

The above analysis demonstrates that two qubits per repeater node are sufficient for fully fault tolerant long distance quantum communication.  To illustrate the possibilities opened up by reduced physical requirements, 
we now turn to a specific example for implementation of
the 
protocol in a solid state system: the nitrogen vacancy (NV) center in diamond.  The qubits required for entanglement connection and purification are realized in the electronic triplet ground state and the nuclear spin of a nearby $^{13}$C impurity.   Each spin can be manipulated by magnetic resonance techniques, and strong hyperfine interactions 
couple the two qubits, allowing experimental demonstration of two-qubit gates~\cite{jelezko04b}.   
A single NV-center
with a nearby $^{13}$C can therefore effectively 
constitute a two qubit quantum computer. 

\begin{figure}\vspace{.0in}
\centerline {
\includegraphics[width=3.5in]{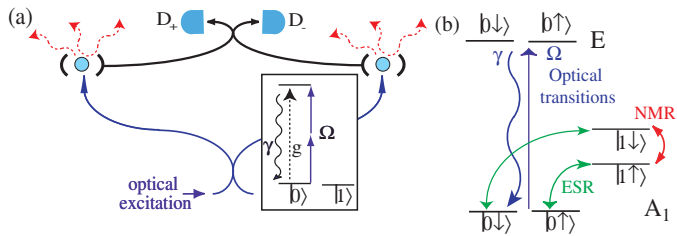}
}
\vspace{.0in}\caption{  (a) Interferometric arrangement for entanglement
generation.  Inset shows relevant level scheme.  (b) Implementation with NV centers.  Electronic spin ($\ket{0}, \ket{1}$) and $^{13}$C nuclear spin
($\ket{\uparrow}, \ket{\downarrow}$) states 
are coupled by  optical, microwave, and RF transitions.  The $M_s = -1$ electronic state can be shifted out of resonance by a small magnetic field.  
} 
\label{levels}\end{figure}

The remaining requirement for a quantum repeater is an entanglement generation scheme. 
In atoms and ions, 
entanglement between spatially separated systems can be 
generated probabilistically by, 
 e.g.,  Raman
scattering~\cite{cabrillo99} or polarization-dependent
fluorescence~\cite{blinov04} 
followed by photon interference.   In the spirit of reducing the physical requirements on the system, we will show that probabilistic entanglement generation can succeed even in the absence of polarization selection rules, allowed Raman transitions, or even radiatively broadened optical transitions.    Our scheme requires only state-selective light scattering (see Fig.~\ref{levels}a, inset), and is thus applicable to a variety of solid-state emitters including the NV center~\cite{weinfurter00,beveratos02, jelezko04}.   
 Furthermore, this simple level structure facilitates entanglement of one qubit (the electron spin) while leaving the other qubit (the nuclear spin) undisturbed.   In particular, by choosing a scattering transition between $M_s = 0$ electron spin states, we can eliminate sensitivity to the nuclear spin state during entanglement generation (see Fig.~\ref{levels}b).  The desire for a simple requirement for level structure combined with the necessity of preserving the nuclear spin state motivates discussion of a new entanglement generation scheme based on state-selective elastic light scattering and photon interference.

Our entanglement generation scheme proceeds as follows:  We consider
two NV centers separated by a distance $L_0$, such that 
each node scatters light only if its  electron spin is in state
$\ket{0}$. 
 Two adjacent nodes thus form state-selective mirrors in
an interferometer  (see Fig.~3(a)).    Our scheme relies on balancing
this interferometer so that when both nodes are in the 
scattering state $\ket{0}$, the outgoing photons will always exit one
detector arm $D_+$.   A detection event in the other arm
$D_-$ 
can then project the spins onto an entangled state. 

We now address this scheme quantitatively, and determine the entanglement fidelity it can produce.  The scheme starts with each node 
in 
the state $(|0\ra +|1\ra)/\sqrt{2}$; state $|0\ra$ is then coupled to an excited level that decays radiatively at a rate $\gamma$.
In the weak excitation limit, we can adiabatically eliminate the
excited state, resulting in coherent scattered light.  
The combined state of node $i$ and the scattered light field is given by  
$\ket{\psi}_i \approx ( \ket{1} + T_i \ket{0} ) / \sqrt{2}$ with  
$ T_i = e^{-\sqrt{P_{\rm{em}}} (\sqrt{1-\epsilon}\hat b^{\dag}_i +
\sqrt{\epsilon} \hat a^{\dag}_i)-P_{\rm{em}}/2}, 
\label{coherentstate}$  
where $P_{\rm{em}}$ is the total emission probability, $\epsilon$ is
the total collection, propagation, and detection efficiency, and $\hat
a_i, \hat b_i$ 
are the annihilation operators for the 
field reaching the beamsplitter and other (loss) fields, respectively.   
In the limit $P_{\rm{em}}\rightarrow 0 $,  a detection event in 
$D_-$ (mode $\hat d_-
\propto \hat a_1 - \hat a_2$) projects the system onto a maximally
entangled state
$\hat d_- (T_1 \ket{01} + T_2 \ket{10})/2 \propto
\ket{\Psi_-}$.
For finite $P_{\rm{em}}$,  there is a chance $\sim P_{\rm{em}}$ 
that 
an undetected photon was emitted into the environment. After a
detection in $D_-$, the nodes cannot be in the $|00\ra$ or $|11\ra$
state; the additional photon emission thus introduces some admixture of the state $\ket{\Psi_+}$ 
(a phase error).   
We find that the 
scheme succeeds with probability 
$P = (1/2)\left(1-e^{-P_{\rm{em}}\epsilon/2}\right)\approx \epsilon P_{\rm{em}}/4$,
producing the state $|\Psi_-\rangle$ with fidelity
\beq
F_0 = \frac{1}{2}\left(1+e^{-P_{\rm{em}}\left(1-\epsilon\right)}\right) \approx 1 - \frac{P_{\rm em} (1- \epsilon)}{2}
\label{fidelity}
\eeq
in time $T_0 \approx (t_0+t_c) /P$.

Finally, we mention 
some technical aspects of the proposed implementation. 
First, 
interferometer stabilization poses a challenge, but has been achieved over $\sim10$ km  distances \cite{holman05}.  
Alternatively the interferometric setup
may be replaced by a photon coincidence detection, which is less
susceptible to path length fluctuations \cite{barrett04,simon}.
Another important source of error is the homogeneous broadening 
typically found in solid-state emitters. 
The effect of this broadening can, however, be reduced by sending the
light through a narrow frequency filter or a using a cavity \cite{andersXX}.
For NV centers coupled to cavities with Purcell factors  $\sim
10$~\cite{santori02}, we find that the dominant source of error is
electron spin 
decoherence. 
Using an emission probability $P_{\rm{em}}\sim 5\%$, a collection efficiency
$\epsilon \sim 0.2$, 
and $t_c \sim70\mu$s over $L_0 \sim 20$ km, we find 
$F_0 \sim 97\%$ for electron spin coherence
times in the range of a few milliseconds.  According to our numerical calculations, this fidelity is sufficient for long distance quantum communication.

In conclusion, we have shown that a fully fault tolerant quantum repeater can be constructed using only two qubits per node.  This opens up the possibility to build repeaters in simple systems with only two degrees of freedom, such as coupled nuclear and electronic spins.  We have exemplified this with a particular implementation in NV-centers, but the concept can be applied to a variety of physical systems \cite{andersXX}. 

     The authors wish to thank P. R. Hemmer, A. Mukherjee,      A. Zibrov, and G. Dutt.  This work is supported by DARPA, NSF, ARO-MURI, and the Packard, Sloan and Hertz Foundations and  the Danish Natural Science Research Council.  


\end{document}